\documentclass[11pt]{article}

\begin{document}
\twocolumn[ \section*{\footnotesize{\it   Spacetime \& Substance}
{\bf 2}, {\rm no. 4, 169-170 (2001) }}

\bigskip

\section*{\small \bf ON THE MASS OF ELEMENTARY CARRIERS OF
GRAVITATIONAL INTERACTION}

\begin{center}
{\bf Volodymyr Krasnoholovets}
\end{center}

\begin{center}
{Institute of Physics, National Academy of Sciences \\ Prospect
Nauky 46,   UA-03028 Ky\"{\i}v, Ukraine \ \
http://inerton.cjb.net}
\end{center}

{ \small {\bf Abstract:} Based on the theory of submicroscopic
quantum mechanics recently constructed by the author the mass of
elementary spatial excitations called inertons, which accompany a
moving particle, is estimated herein. These excitations are
treated as carriers of both inertial and gravitational properties
of the particle.
\\

{\bf  Key words:} \ \  space, inertons, mass, gravitation
\\

{\bf PACS:} \ \ 03.65.Bz Foundations, theory of measurement,
miscellaneous theories. \\  03.75.-b Matter waves. \ \ 04.60.-m
Quantum gravity
\\
\medskip \vspace{1mm} } ]

\hspace*{\parindent}

In a series of recent papers by the author [1-4] submicroscopic
quantum mechanics has been constructed. It easily results in the
Schr\"odinger and Dirac formalisms on the atom scale. The theory
has predicted the existence of peculiar spatial excitations around
a moving particle. Those excitations have been identified with a
substructure of the matter waves and been called "inertons". The
theory has successfully been verified experimentally, namely, it
has been demonstrated how inertons manifest themselves in numerous
experiments [5] and, moreover, inertons have been recorded in our
experiments as well [6-8].

   Detailed theoretical consideration [1-5] of the motion of a
canonical elementary particle in the real space that is
characterized by a submicroscopic structure (and the particle is
an element of the space as well), allows an evaluation of the
value of mass of elementary excitations -- inertons -- of a space
substrate (i.e. quantum aether).

Elementary excitations of the space substrate are created due to
collisions of a moving particle by superparticles -- primary cells
of the structure of the real space. The excitations were called
inertons [1] because, in essence, they reflex inert properties of
the particle, i.e. inertons appear owing to the resistance which
the particle experiences at its motion on the side of the space
substrate that in turns is specified by quantum properties.
Amplitude $\Lambda$ of the particle's inerton cloud that surrounds
the moving particle obeys the relationship [1]
\begin{equation}\label{1}
  \Lambda= \lambda {\kern 1pt} {c}/{v_{\kern 1pt 0}}
\end{equation}
where $\lambda$ is the de Broglie wavelength of the particle,
$v_0$ is its initial velocity, and $c$ is the velocity of light.

According to the author's concept [1-5], a particle is created
from a degenerate superparticle. Therefore a particle should be
treated as a local curvature of the space. The notion of the mass
is associated with the alteration of an initial volume of the
mother superparticle. Around a particle, a deformation coat, or
crystallite, is formed that consists of superparticles which
possess mass. Beyond the crystallite superparticles are massless.
Thus the crystallite plays the role of a screen that shields the
particle from the degenerate space substrate. The total mass of
massive superparticles of the crystallite is equal to the mass
$m_{v_{_0}}$ of the particle [4], which is found in the center of
the crystallite. The size of the crystallite is estimated by the
length of the Compton wavelength ${\tilde \lambda}_{v_{_0}}=
h/m_{v_{_0}} c$. As the solid state physics teaches, the
availability of the crystal structure automatically implied the
appearance of elementary vibrational excitations in the
crystallite [4].

  Inertons should be considered as a substructure of the matter
waves [3-5]. In papers [1,2,4] it has been noted that the kinetic
energy of an emitted inerton is directly proportional to the
energy of the particle that the particle had had at the moment of
the collision with the vibrating mode of the crystallite. Both the
energy of the particle and that of the mode decrease from
collision to collision. Therefore the same should occur for
emitted inertons: the energy of the $(i+1)$th inerton is less than
the energy of the $i$the one. Since we assume that the initial
velocity of emitted inertons has an order of the velocity of light
$c$, we should conclude that the mass of emitted inertons
gradually decreases as well, i.e. $m_{i+1} < m_i$.

   For instance, if the velocity of the particle $v_{\kern 1pt 0} \ll c$, then
the inequality $m_i> m_{\rm cr}$ holds in the beginning, where
$m_{\rm cr}$ is the averaged mass of the crystallite's
superparticle. In other words the inequality is correct at a small
value of $i$ (we recall that $i=\overline{1, N}$ where $N$ is the
total number of collisions of the particle along its half de
Broglie wavelength $\lambda/2$. In this case amplitude $\Lambda_i$
of the $i$th emitted inerton prevails the crystallite size,
$\Lambda_i \gg {\tilde\lambda}_{\kern 1pt v_{_0}}$.

However, as the index $i$ increases, the inerton mass diminishes
and reaches values less than the mass of crystallite's
superparticles, $m_i<m_{\rm cr}$. In this case the amplitude of
the inerton has a magnitude under the crystallite size,
$\Lambda_i<{\tilde \lambda}_{\kern 1pt v_{_0}}$.

The averaged mass $m_{\rm cr}$ of a superparticle in the
crystallite can be estimated. For example, in the case of a
nonrelativistic electron the Compton wavelength ${\tilde
\lambda}_{\kern 1pt 0}=2.42\times 10^{-10}$ cm. If we divide the
crystallite volume ${\tilde\lambda}_{\kern 1pt 0}^{\kern 1pt 3}$
by the volume of a superparticle ${\cal V}\sim (10^{-28})^3$
cm$^3$, we will get the number of superparticles in the
crystallite $N_{\rm in \ crys.}\sim 10^{55}$. Since the mass of
the crystallite is taken to be the mass of the particle, we will
obtain the following value for the mass of a crystallite's
superparticle: $m_{\rm cr}= M_0^{\rm electron}/N_{\rm in \
crys.}\sim 10^{-85}$  kg.

We can also evaluate the mean mass $\bar m_{\rm in}$ of an emitted
inerton. For this purpose we should divide the total mass $\Delta
M= (M_0/\sqrt{1-v_{\kern 1pt 0}^2/c^2}-M_0)$  of the emitted
inerton cloud by the number of emitted inertons $N=\lambda/{\cal
V}^{1/3}$ (we recall that the cloud is emitted along the first
half de Broglie wavelength $\lambda/2$, then it is absorbed in the
next section $\lambda/2$ of the particle path, and so on).

Setting $v_{\kern 1pt 0}$ equals $0.01 {\kern 1 pt} c$ to $0.999
{\kern 1pt} c$, we obtain:
\begin{eqnarray}\label{2}
\bar m_{\rm in} &=& \frac {{M_0}/{\sqrt{1-v_0^2/c^2}}- M_0 }
{\lambda/{\cal V}^{1/3}}             \nonumber         \\ &=&
10^{-57} \ {\rm to} \ 10^{-45} \ {\rm kg}.
\end{eqnarray}

In the case of inertons emitted by atoms which vibrate in a solid,
$\bar m_{\rm in}$ falls in the broad range between values (2) and
about $\sim 10^{-70}$ kg (at the extremely low atom velocity $v_0
\sim 1$ $\mu$m/s).

The value of mass of carriers of a peculiar interaction between
objects was estimated also by other researchers. For instance,
Kolpakov [9] studying experimentally a nonelectromagnetic
interaction between both extrasensitive participants and objects
of abiotic environment proposed a pure classical mechanism of the
propagation of excitations of an aether substance; his model
yielded evaluation $\sim 10^{-73}$ kg for the mass of carriers of
the revealed interaction.

Starting from the field formulation of the general theory of
relativity Zhuk [10] obtained for his "gravitons",  carriers of
the gravitational interaction, mass $\sim 10^{-69}$ kg. It is
interesting that this magnitude is approximately equal to an
average value between mentioned inerton masses $m_{\rm cr}$ and
$\bar m_{\rm in}$.

Thus, it can be concluded based on submicroscopic quantum
mechanics that the value of mass of carriers of the
inertial/garvitational interaction is not strongly fixed. Masses
of inertons emitted and then absorbed by a moving particle is
distributed in a wide spectral range. Note that a similar
situation occurs in the case of the electromagnetic radiation: the
photon frequency can vary from practically zero to the frequency
of high-level $\gamma$-photon.

\end{document}